\documentclass[12pt]{article}

\input{tcilatex}

\begin{document}

\author{S. S. e Costa and H. V. Fagundes \\
Instituto de F\'{i}sica Te\'{o}rica, Universidade Estadual Paulista,\\
Rua Pamplona, 145, S\~{a}o Paulo, SP 01405-900, Brazil,\\
E-mail: sancosta@ift.unesp.br, helio@ift.unesp.br}
\title{On the birth of a closed hyperbolic universe}
\maketitle

\begin{abstract}
We clarify and develop the results of a previous paper on the birth of a
closed universe of negative spatial curvature and multiply connected
topology. In particular we discuss the initial instanton and the second
topology change in more detail. This is followed by a short discussion of
the results.

\end{abstract}

\section{Introduction}

In a recent paper \cite{eCF} we\footnote{%
In Ref. \cite{eCF} the first author appeared by mistake with name S. S. da
Costa. His correct name is S. S. e Costa, as above.} suggested a process for
the spontaneous creation of a universe with closed - i. e., compact and
boundless - spatial sections of negative curvature. (A short report on the
same subject was presented at the \textit{Cosmological Topology in Paris
1998 }meeting \cite{CTP98}.) This process involved four steps: (i) the
actualization of an instanton of nontrivial global topology into a de Sitter
universe of positive spatial curvature; (ii) a topology and metric change
into a closed de Sitter world of negative spatial curvature; (iii) inflation
of this hyperbolic de Sitter universe; and (iv) reheating and beginning of
the radiation era with the metric of Friedmann's open model $(\Omega
_{0}<1,\ \Lambda =0)$ and the spatially compact topology obtained in step
(ii). In Sections 2 and 3 we justify and develop steps (i) and (ii) in more
detail. Steps (iii) and (iv) may be taken as the same as in the usual
inflationary scenarios - see \cite{KT}, Chapter 8, for example. The last
section briefly argues for the compatibility of a compact hyperbolic
universe both with the observed fluctuations of the cosmic microwave
background (CMB) and with an inflationary scenario leading to a present
density ratio $\Omega _{0}<1$.

\section{The instanton orbifold}

We modeled the spontaneous birth in Vilenkin \cite{Vil}. But while he has an 
$S^{4}$ instanton tunneling into an $R\times S^{3}$ spherical universe
(where $S^{n}$ is the $n$-sphere and $R$ is the time axis), we start from a
more complex structure in order to reach a spherical spacetime $\mathcal{M}%
_{L}=R\times (S^{3}/\Gamma )$ with nontrivial topology. Here $M=$ $%
S^{3}/\Gamma $ is the quotient space of $\ S^{3}$ by a discrete, finite
group of isometries$\ \Gamma $, which acts freely on $S^{3}$; cf. \cite
{Scott}, for example. If $S^{3}$ has unit radius the volume of $M$ is $2\pi
^{2}/|\Gamma |$, where $|\Gamma |$ is the number of elements of $\Gamma ,$
so we have a variety of spherical manifolds that may, in principle, be
chosen as spatial sections of positive curvature for a Robertson-Walker
model. In the example of \cite{eCF} $M$ is the lens space $L(50,1),$ with
volume $2\pi ^{2}/50.$

Instead of $S^{4}$ we construct a more general instanton $S^{4}/\Gamma $,
which we proceed to describe. The action of $\ \Gamma $ on $%
S^{4}=\{(X_{\alpha },\alpha =0-4)\in R^{5};\ X_{\alpha }X_{\alpha }=1\}$ is
obtained by extending its action on the standard (unit radius) $S^{3}$ to
all `parallel' 3-spheres on $S^{4}$, that is, for $|X_{0}|$\textbf{\ }$\leq
1,$ $S_{X_{0}}^{3}=\{(X_{0},\ X_{i},i=1-4);\ X_{i}X_{i}=1-X_{0}^{2}\}.$ The
action is already defined on the `equator' $S_{0}^{3},$ which is isometric
to $S^{3}.$ Let $(X_{0},X_{i})$ $\in S_{X_{0}}^{3}$ and $\gamma \in \Gamma .$
If $\ |X_{0}|$\textbf{\ }$<1,$ then $(0,\ X_{i}^{\prime }=X_{i}/\sqrt{%
1-X_{0}^{2}})\in S_{0}^{3},$ so that $\gamma (0,X_{i}^{\prime
})=(0,X_{i}^{\prime \prime })\in S_{0}^{3},$ and we define $\gamma
(X_{0},X_{i})\equiv (X_{0},\ X_{i}^{\prime \prime }\sqrt{1-X_{0}^{2}})\in
S_{X_{0}}^{3}.$ If $|X_{0}|=1$, then $\gamma S_{\pm 1}^{3}=S_{\pm 1}^{3},$
which are the poles of $S^{4}.$ Thus the action of $\ \Gamma $ on $S^{4}$
is not free, and so the quotient space $S^{4}/\Gamma $ is not a manifold, but
an orbifold with two cone points corresponding to the poles of $S^{4}$ - cf.
Scott \cite{Scott}, Sec. 2.

Actually only the lower half ($X_{0}\leq 0)$ of the instanton takes part in
the solution. Following Gibbons \cite{Gibbons} we call this manifold $%
\mathcal{M}_{R}$ - the index $R$ meaning Riemannian (the positive definite
part of the solution, popularly known as Euclidean on account of the metric
signature). The full spacetime solution is $\mathcal{M=M}_{R}\cup _{\Sigma }%
\mathcal{M}_{L},$ where $\mathcal{M}_{R}$ and $\mathcal{M}_{L}$ are attached
smoothly by $\Sigma =S_{0}^{3}/\Gamma =\partial \mathcal{M}_{R}.$ With this
generalization Gibbons's conditions are satisfied: $\mathcal{M}_{R}$ is a
compact orbifold with $\Sigma $ as sole boundary; $\Sigma $ is a Cauchy
surface for $\mathcal{M}_{L}$; and it has a vanishing second fundamental
form with respect to both $\mathcal{M}_{R}$ and $\mathcal{M}_{L}$ - this is
true of the $S^{3}$ covering, and the action of $\Gamma $ does not interfere
with the local metrics.

\section{The second topology change}

As described in \cite{eCF} the first epoch after creation had the metric

\begin{equation}
ds^{2}=-dt^{2}+r_{0}^{2}\cosh ^{2}(t/r_{0})(d\chi ^{2}+\sin ^{2}\chi \
d\Omega ^{2})\ ,  \label{deSitterS}
\end{equation}
where $d\Omega ^{2}=d\theta ^{2}+\sin ^{2}\theta \ d\varphi ^{2}$ and $r_{0}$
is Planck's length or time; and the topology $R\times M$ discussed in the
preceding section. Then we assumed a formalism developed by De Lorenci et
al. (\cite{DLor}; hereafter LMPS) could be used to justify a quantum
transition into a second epoch with topology $R\times M^{\prime },$ where $%
M^{\prime }$ is a compact hyperbolic manifold, and metric 
\begin{equation}
ds^{2}=-d\tau ^{2}+r_{0}^{2}\sinh ^{2}(\tau /r_{0})({d\chi ^{\prime }}%
^{2}+\sinh ^{2}\chi ^{\prime }\ d\Omega ^{2})\ ,  \label{deSitterH}
\end{equation}
In the example of \cite{eCF} $M^{\prime }$ is Weeks manifold, which is the
smallest space in the \textit{SnapPea} census \cite{JW}.

To match these two stages we postulated conservation of physical volume. But
in order to use the results in LMPS we should rather have continuity of the
expansion factor: if $t_{f}$ is the final time of stage one and $\tau _{i}$
is the initial time of stage two, then this continuity requires $\cosh
(t_{f}/r_{0})=\sinh (\tau _{i}/r_{0})$. The homogenizing process to be
produced by inflation in stage two demanded that $\tau _{i}$ was of the
order of Planck's time $r_{0}.$ To keep a number from the example in \cite
{eCF}, let $\tau _{i}=0.9865r_{0}.$ If follows that $t_{f}/r_{0}=0.5489$. In
that example this time interval would not allow for the homogenization of
space $M$. However, this first stage is so short that it may eventually, in
a complete theory, be viewed as a quantum intermediate state. Anyway, it
probably does not make sense to speak of density smoothening in a
sub-Planckian scale. As for the universe's homogenization, it is taken care
of by the 70-odd $e$-fold inflation of our second epoch, as in more usual
scenarios.

Now we proceed to give estimates of the probabilities for the topology
change between these stages, according to LMPS. \ It would be desirable to
obtain absolute probabilities, but in the present stage this is not
possible, because their wave functions are not normalized. LPMS calculate
conditional probabilities for transitions among three topologies on
manifolds $M_{k},$ one for each sign of the curvature, $k=0,\ \pm 1.$ Here
we shall restrict ourselves to $M$ and $M^{\prime }$; the case for a
Euclidean manifold $M_{0}$ is unclear, given the arbitrariness and
continuous range of its fundamental polyhedron's volume.

We need an additional hypothesis in order to apply LMPS's results. The
latter assumes null potentials $U(\phi )=V(\xi )=0,$ but since these
potentials enter their Hamilton-Jacobi equation only in the combination $%
U(\phi )+V(\xi ),$ the same equation is obtained by only requiring $U(\phi
)=-V(\xi )>0.$ Although this condition looks contrived, we need it at
present because our transition in \cite{eCF} was supposed to take place near
the false vacuum.

In LMPS the calculations hinge on functions $F_{k}$, which we rewrite, in
Planckian units, as

\begin{equation}
F_{k}(M_{k})=\frac{\bar{a}}{2\pi m}\int_{M_{k}}\cos (2\sqrt{k}\chi )\sin
\theta \,d\chi \,d\theta \,d\varphi \ ,  \label{Fk}
\end{equation}
\smallskip where $\bar{a}$ is the expansion factor at the moment of the
transition, and $m$ is the mass associated with an auxiliary field $\xi $,
which ``is introduced to give a notion of time evolution to the quantum
states.'' (This field is their version of Kuch\u{a}r and Torre's \cite{KandT}
``reference fluid.'')

The last equation gives immediately $F_{1}(M)=0,$ because for the lens space
the range of $\chi $ is $[0,\pi ]$ for any values of $\theta \,\ $and $%
\varphi .$

For $k=-1$ Eq.\thinspace (72) in LMPS turned out to be impractical for
actual evaluation; only lower and upper bounds were obtained for their $%
F_{-1}(I^{3}).$ We succeeded in performing the integration in our case by
first expressing Eq. (3) in hyperbolic cylindrical coordinates $(\rho
,\varphi ,z),$ which are related to the spherical coordinates $(\chi ,\theta
,\varphi )$ by $\sinh \rho =\sinh \chi \sin \theta ,$ $\tanh z=\tanh \chi
\cos \theta ,$ and $\cosh \chi =\cosh \rho \cosh z.\ $Then we get

\begin{equation}
F_{-1}(M^{\prime })=\frac{\bar{a}}{2\pi m}\left[ 2V(M^{\prime
})+\int_{M^{\prime }}\frac{\sinh \rho \cosh \rho \ d\rho \ d\varphi \ dz}{%
\cosh ^{2}\rho \cosh ^{2}z-1}\right] \ ,  \label{Fm1}
\end{equation}
where $V(M^{\prime })=0.942707$ is the volume of Weeks manifold. The
integral was calculated by decomposing the fundamental polyhedron for $%
M^{\prime }$ into quadri-rectangular tetrahedra, and using results of
hyperbolic geometry as given by Coxeter \cite{Coxeter} and Coolidge \cite
{Coolidge}. This computation was carried out by one of us (SSC) , and is
discussed elsewhere \cite{SSC}. The result is $F_{-1}(M^{\prime })=1.4777\ 
\bar{a}/m.$

Let the wave function of the universe be $\Psi (a,\phi ,\xi ,M_{k})$, where $%
a $ is the expansion factor and $\phi $ is the inflaton field. Similarly to
LMPS we put $|\Psi (\bar{a},\bar{\phi},\xi ,M^{\prime })|^{2}=A(\bar{a},\bar{%
\phi})\exp (2F_{-1}\xi ),\quad $ $|\Psi (\bar{a},\bar{\phi},\xi ,M)|^{2}=C(%
\bar{a},\bar{\phi})\exp (2F_{1}\xi ),$ where $A$ and $C$ are positive
functions. Then the ratio of probabilities that the universe is found with
spaces $M^{\prime }$ and $M$ at ``time'' $\xi $ is $P(M^{\prime
})/P(M)=(A/C)\exp (2.9554\,\bar{a}\xi /m).$ This is null for $\xi =-\infty ,$
which implies initial space $M,$ and infinite for $\xi =+\infty ,$ hence
final state $M^{\prime }.$ Thus we get the desired topology change.

We are aware that LMPS's formalism suffers from the usual doubts and
limitations of quantum cosmology calculations. But we hope it is a step in
the right direction.

\section{Discussion}

Recently the theoretical preference for flat space cosmology has been
reinforced by observations - see, e. g., \cite{lambda} and references there
- that suggest a substantial present value of the cosmological constant $%
\Lambda ,$ making up a total critical density: $\Omega _{0}=\Omega
_{matter}+\Omega _{\Lambda }=1.$ But this belief is not universal - cf. \cite
{Aguirre}, for example; should it become untenable, we may have to face a
subcritical density and a universe with negative spatial curvature. There is
even the possibillity of $\Omega _{0}$ $<1$ in the presence of a positive $%
\Omega _{\Lambda }$; cf. Quast and Helbig \cite{QH} and references there.
Recent observational results, as quoted by Lehoucq et al. \cite{LeUzLu},
only restricts $\Omega _{0}$ to the range $[0.88,1.12]$. 

It has been argued \cite{Pog} that the CMB fluctuations are incompatible
with a closed hyperbolic model (with $\Lambda =0$) unless $\Omega _{0}$ $%
\approx 1,$ and its spatial dimensions are of the order of magnitude of the
observable universe. The recent work of Aurich \cite{Aurich} seems to
contradict this. See also Inoue et al. \cite{Inoue}, Cornish and Spergel 
\cite{CS}. But even if Bond et al. \cite{Pog} are correct, the case for a
closed hyperbolic universe still deserves investigation. And since it might
not be small enough \cite{ES86} to account for the homogeneity of cosmic
images (the substitute for the true homogeneity of simply connected models),
we should be prepared to associate compactness with inflation, as discussed
in \cite{ES86} and done here. The usual inflationary scenario tends to
exclude the open Friedmann model on the grounds of a needed fine-tuning of
the density ratio $\Omega (t)$ in early times. Thus at the beginning of the
radiation epoch in our model, $t_{1}=71t_{Planck},$ the equations in \cite
{KT}, Chapter 3, indicate $\Omega (t_{1})\approx 1-1\times 10^{-57},\,$which
looks suspicious for the open model. However, if we find that creation and
early evolution were governed by topological constraints, then the fact of a
pre-inflationary negative curvature being diluted by inflation could only
lead to a value of $\Omega (t_{1})$ that was very close to, but still
smaller than one. This is so because the by then frozen topology on a
compact 3-space could not support a Euclidean metric - cf. \cite{Thurston}.
(A similar argument has been made by Padmanabhan \cite{Pad}, but it does
seem to hold in his context of infinite spatial sections.)

\bigskip

One of us (SSeC) thanks Funda\c{c}\~{a}o de Amparo \`{a} Pesquisa do Estado
de S\~{a}o Paulo (FAPESP - Brazil) for a doctorate scholarship.

\end{document}